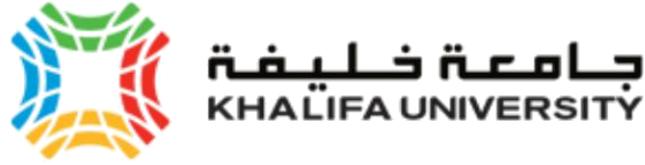

ECCE DEPARTMENT

SENIOR DESIGN PROJECT

# WEARABLE POSTURE MONITORING SYSTEM WITH VIBRATIONAL FEEDBACK

### GROUP MEMBERS

| | |
|---|---|
| Alyazyah Alsuwaidi | 100039053 |
| Aisha Alzarouni | 100038603 |
| Dana Bazazeh | 100038654 |

### ADVISORS:

Dr. Nawaf Almoosa
Dr. Kinda Khalaf
Dr. Raed Shubair

SUBMITTED | DEC, 11, 2016

# ABSTRACT

Around 50 billion dollars is spent yearly on therapy for low back pain in the United States alone. Low back pain is one of the most common reasons for doctor visits. Having poor posture has been found to be a main cause of lower back pain as it impacts the transverses abdominis muscle. Maintaining a good posture and changing one's position from time to time is considered to significantly improve and maintain one's health. The world has witnessed a vast amount of smart monitoring devices that are used to enhance the quality of life by providing different types of support. Smart wearable technology has been the main focus of this century, specifically in the medical field, where the advances range from heartbeat monitors to hearing aids. This report highlights the design, development and validation process of a compact wearable device that uses multiple sensors to measure the back posture of a user in real time and notify them once poor posture is detected.

# TABLE OF CONTENTS

# LIST OF FIGURES

# LIST OF TABLES

# 2. INTRODUCTION

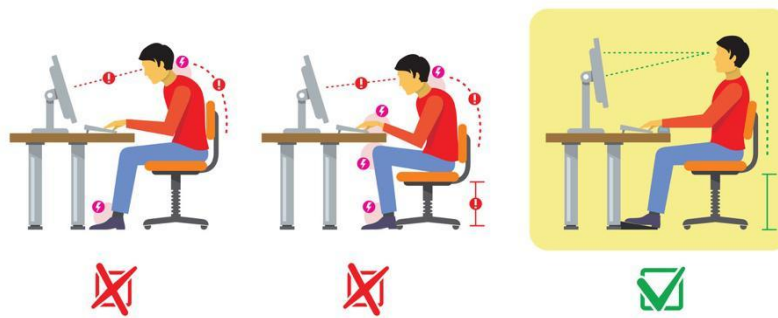

Figure 1: Bad posture effects the whole body

The last two decades have witnessed an exponential growth and tremendous developments in wireless technologies and systems, and their associated applications, such as those reported in [1-23].Posture is the way people carry themselves, in the way they stand, sit, walk and perform tasks, and this posture has a substantial effect on their health. Maintaining a good posture allows the vertebras of the spine to be correctly aligned. Poor posture has been linked with bad health as well as lower performance. A study showed that having a slouched posture impacts the transverses abdominis muscle. It was shown that the thickness of the transverses abdominis muscle is significantly less when a person maintained a slouched posture [25]. This transversus abdominis dysfunction is directly associated with low back pain. Low back pain is one of the leading cause of disability in the world, where it is estimated that around 80% of the world population will experience it at some point in their lives [26]. Each year in the United States, around 50 billion dollars is spent on treating back pain [27]. This problem can also be seen in the UAE where 62% of the young population report to suffer from back pain [28]. Dr. Hilali Noordeen, an orthopedic surgeon at Burjeel Hospital in the UAE said that back pain caused by daily habits such as sitting on a chair with a hunchback, can have severe damaging effects in the long run [28]. Moreover, another study showed that subjects asked to sit with a hunched posture, reported greater stress and thus lower performance [29]. Maintaining the same position for a long time, even with good posture, is also considered a bad postural habit as the muscles in the spine may stop producing substances that are essential for normal biological operations [30]. Therefore, keeping a good posture and changing one's position from time to time is considered important, if not necessary for maintain good health.

## 2.1 PROBLEM STATEMENT

A large number of the working population have office jobs that require long hours of sitting, and students spend most of their time on laptops devices leaning forward and these habits lay significant strain on the neck and back. In fact, after surveying 200 people of various age groups, 76% said that they do not practice good posture. Of those people, more than 90% said that they experience back pain occasionally to constantly.

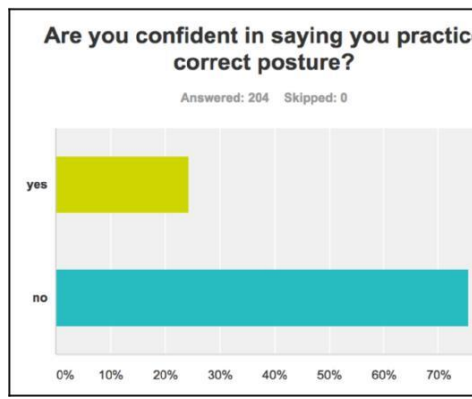
Figure 2: survey question 1

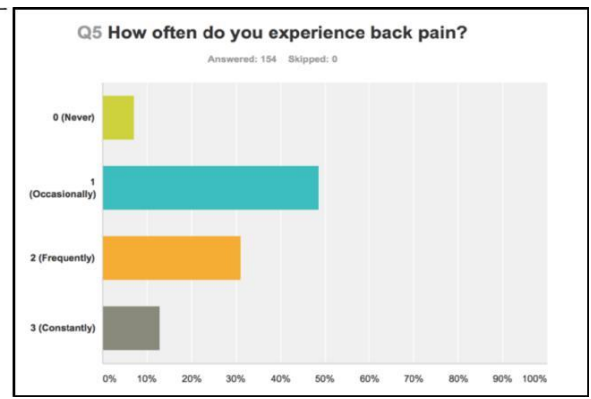
Figure 3: survey question 2

This unintentional repetition of poor posture every day slowly changes the body's structure and the body ends up adapting to it. Bad postural habits cause the chest muscles to tighten leading to an excessively curved back in the upper back or thoracic region as shown in figure 4 below. The muscles of the upper back loosen and eventually weaken. Therefore, we introduce a smart monitoring system that is used to enhance the quality of life by providing the support needed to maintain good posture and keep the body moving.

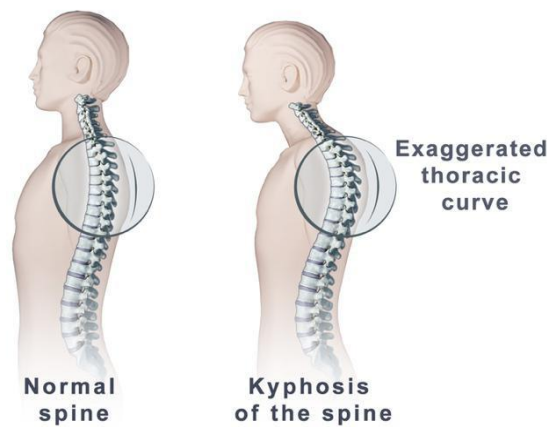
Figure 4: Normal Spine versus Kyphosis of the spine [31]

The system will consist of a wireless wearable device that can be attached to a person's back and is used to alert the user in case of an unhealthy postural situations, such as sitting with a hunchback or remaining idle for long periods of time. By motivating the user to exercise posture control, we can help improve the user's ability to maintain an upright position of the spine on the long term with no external help, as well as reduce the possibility of future musculoskeletal problems and pain. The user will acquire awareness of their own posture using vibrational feedback, and correct it when necessary. With time, the device will unconsciously engrave the habit of maintaining a good posture, without the need for the user to be reminded. Such that, the user will be able to fix their own posture after continued use. A mobile app will also compliment the device, showing graphs of posture over time, and sending the users a summary of their progress of maintaining good postural habits so that they can track their own progress.

## 2.2 OBJECTIVES

### 2.2.1 BENEFITS

- Reduce chronic pain that results from poor posture.
- Decrease the stress and pressure on the spine.
- Prevent musculoskeletal disorders and structural deformity of spine.
- Train users to maintain good back posture until it becomes a daily routine.

### 2.2.2 FEATURES

- Compact and light.
- Wireless connectivity.
- Calibrated measurements to various users.
- Rechargeable battery.

### 2.2.3 GOALS

To reduce the repercussions of bad posture by designing a wearable device that detects and corrects poor posture via haptic feedback, and trains users to maintain posture through continuous use.

### 2.2.4 FUNCTIONS

- Collecting data from sensors positioned on the thoracic curve.
- Processing the data obtained by converting them to angular values.
- Sending feedback to the user via both vibration motors to alert him when poor posture is detected, and an App to track user's posture during the day.

## 2.3 LITERATURE REVIEW AND EXISTING TECHNOLOGIES

Despite widespread acceptance that work related spinal disorders and back pain can be prevented by fixing the posture through ergonomic solutions such as adjustable chairs or other modifications applied to the workstation [32], There exists very few reliable and accurate methods of continuously monitoring posture in any environment for postural modification. Recently, wearable sensing technologies targeting posture monitoring have been attracting increasing attention, making postural monitoring not restricted to a single location, thus enabling real time measurements and monitoring of posture. Different solutions for monitoring postural activity are reported in the literature, researchers working on such technologies are attempting to integrate wearable computing devices and sensors into textiles for data collection. However, due to the novelty of this field and unavailability of datasets and information on sensor development done by previous researches, most effort is still directed towards the development of the measuring method and accuracy. Thus, the integration of the sensors into a complete system as well as aspects like comfort, aesthetics and wearability are often neglected. Previous work can be classified into three categories in term of the adopted method of obtaining measurements:

First category of work is based on fiber optic sensors. These sensors use the intensity of light passing through as a measurement tool which is proportional to the bend in the sensor. Dunne et al. built up a system that relates the optical fiber sensor readings to the user's sitting posture [33]. Fiber optic sensors were also used to measure seated spinal posture in

[34]. However, this solution is limited to seated bending back postures. Second category of work turns to make use of non-intrusive pressure sensors. Dunne et al. [35] used textile piezo-resistive pressure sensors to measure shoulder and neck movements through registering the pressure between skin and textile. However no test was performed to detect the sensors reaction to movements of varying magnitudes. Third category of works explores the use of

accelerometers. Hanson et. al. [13] employed the use of accelerometers to measure joint angles with good accuracy results. Similarly, Van Laerhoven et al. [37] and Martin et al. measured postural activities using accelerometers by attaching the to the pants. Lin et al. [15] presented a multi-posture monitoring system using accelerometers embedded in a wearable vest. Moreover, a commercially available product, lumo lift [38], also helps monitor and coach upper body movements through the use of a gyroscope. The emphasis of this product however, is on the aesthetic integration of posture feedback to the wearable device rather than on reliable measurements. Some of the mentioned techniques above are not well suited to develop a wearable measuring system due to the weight of the sensor or to the impracticality of placing it on the user's back.

Our approach is similar to the postural monitoring system developed by Wang et al.[40] Where two sensors attached to a garment are used in reference to each other to measure the spinal angle. However, our approach is more automated as their system relied on a therapist calibrating the system on each user to a range of various postures. Moreover, because the measurements in the postural monitoring system is dynamic, accelerometers can not give the same accuracy as in the static case. Thus, in addition to the inertial acceleration, the gravitation must also be considered. Accelerometers and gyroscopes are commonly incorporated for dynamic positional changes. To measure inclination in spinal curvature, angular rates of rotations are integrated to obtain the positions in the sagittal plane. However, the position values are still due to some inaccuracies as they are affected by drift problems due to the integration method used [41]. On the other hand, the AHRS modules have an accuracy of better than 1 degree, and have high accuracy in dynamic estimations, as they use a combination of accelerometer, gyroscope and magnetometer for stabilization in measurements making them a better option for data collection. The next section in this report will discuss the AHRS and will provide more technical information on the accuracy and data collection method.

# 3. REQUIRMENTS AND SPECIFICATIONS

After taking into account all the existing technologies and studies performed in this area as discussed in the literature review section, a detailed description of our design is presented in the following section. This section will discuss all the requirements our system will be based on, the theory behind the data collection and obtaining a measure for detecting postural changes, and the various components and technologies the system will encompass. We will talk about the main requirements of our design, concept of Euler and quaternion angles and how the thoracic angle is obtained through them. Moreover, we discuss the several hardware components and their specifications and we portray the iterative design and concept generation process, shedding light on how we came up with our final design structure.

## 3.1 WEARABLE SYSTEM REQUIREMENTS

The hierarchy chart below summarizes the requirements of the postural monitoring system. The following requirements should be followed when coming up with design alternatives for the system. A more detailed description of each requirement follows below the chart.

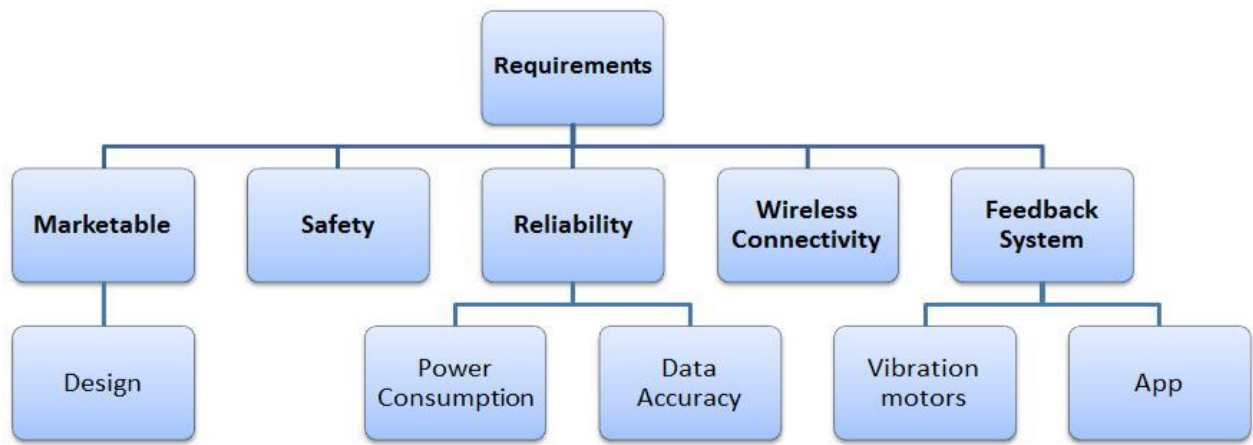

Figure 5: System requirements

- **Marketable**

The design of the system is one of the most important factors in wearables. For example, Size and weight are very critical. The design should be small, light and comfortable to the user to be worn under the shirt conveniently.

- **Safety**

Safety is highly important as well. The system must not heat up especially when it's placed directly on the skin. The design shouldn't also cause skin reactions.

- **Reliability**

The system should be reliable in terms of power consumption and data accuracy. Since wearable devices are powered by batteries and are required to be ON almost all the time, they consume a lot of power. Reducing their power consumption is considered challenging because the capacity of the battery is limited since the overall size has to be small. Therefore, wearable should operate at ultra-low power. If Bluetooth is intended to be used, then Bluetooth Low Energy is the best choice for reducing power consumption.

In order to insure data accuracy, an IMU and a flex sensor will be used to to determine the user's posture instead of one sensor. Additionally, the sensor will be calibrated to meet different heights and body shapes for accurate results.

- **Wireless connectivity**

Wireless connectivity is significant in wearables especially when there is an output to display. The system should be able to communicate with a user interface to display the collected data and interact with the user.

- **Feedback System**

The system should provide feedback to the user via haptic. In system's design, a vibration motor will be placed in order to alert the user (with vibrations) when poor posture is detected. Vibration motors should be energy efficient, small and flat.

## 3.1 MODULES REQUIREMENTS

Following the wearable system requirements, below is a detailed description of each of the modules that the system will include. Some modules were added after the first prototype due to testing with various sensors and components. Moreover, each module or component has a verification method in order to make sure that the requirements are reached.

Table 1: Requirements of modules

| MODULE | REQUIREMENTS | VERIFICATION |
|---|---|---|
| **IMU** | Data acquisition and angles should be accurate along x,y,z axis | The IMU will be connected to the microcontroller and rotated along the axis while obtaining data in real time. |
| **Flex Sensor** | The bending of the flex sensor should be accurately measured along the horizontal axis of motion. | Check the detection of back slouching and bending through an MCU while placed on a person. |
| **Vibration Motor** | Buzzes for a specific period of time only (while user is slouching) and stops vibrating when good posture is maintained. | Connect it to Arduino board through an output digital pin and apply slouching case with a period of time set in the program to send a high signal to the motor. |
| **Wireless Module** | Bluetooth modem should consume a low energy and should efficiently send data to a user interface. | Real time data sent to a user interface will be formed as a graph to allow the user view his performance during a day.<br><br>A low energy Bluetooth module will be used to avoid consuming a lot of energy. |
| **Battery** | The voltage supplied should be greater than 3V, and device it shouldn't heat up. | Measure the power consumption of each component to ensure good power consumption for a wearable device.<br><br>Place the battery on the back and test if it heats up. |

## 3.3 DESIGN ALTERNATIVES

Following the design and module requirements, we come up with different design alternatives for the wearable device. The figure below demonstrates the design alternatives that the team has come up with. Each design has a set of pros and cons based on the requirements of the system that were discussed in the sections above.

In the first design cycle, we looked at the different ways our postural monitoring system can be implemented. These ways could target different parts of the body to detect bad postures. Each design option is further explained below.

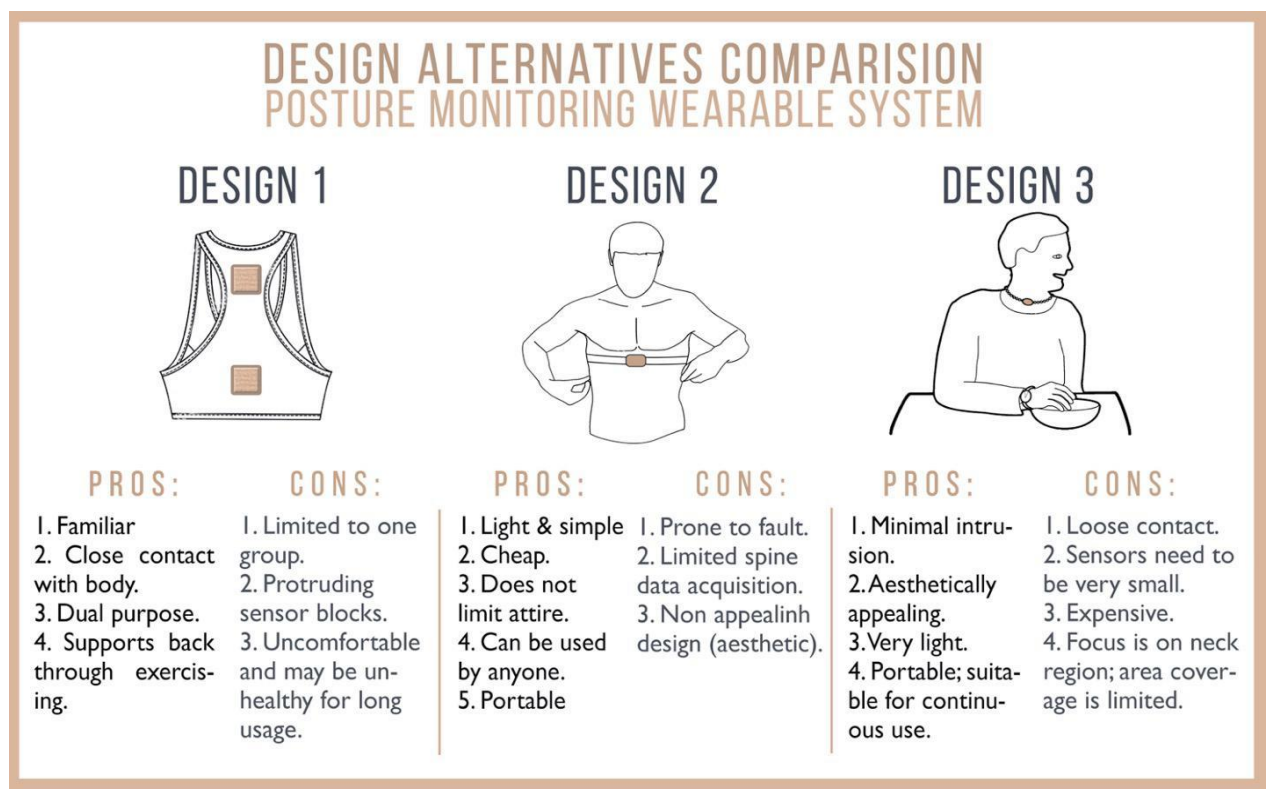

Figure 6: Design alternatives

1) **Undershirt**

An under-shirt, as depicted in figure 4, covers a large area of the spine and gives us more flexibility with sensor placement. However, people might not be comfortable wearing it throughout the day as it needs to be tight around the body in order for the sensor to collect correct angle measurements and there might be protruding parts. Moreover, sizing might be an issue to people of various heights as sensor placements may differ.

2) **Back strap**

The strap can be discreet since it can be worn under clothes, and can be worn everyday. It is light and comfortable and allows us to use a muscle flex sensor to measure muscle strain and detect stressed muscles which is also an issue related to posture. However, a downside to this implementation is how the accelerometer can be used to detect degree of bending to detect bad posture.

3) **Necklace**

Unlike the above options, this one does not target the back directly. It aims to detect posture by monitoring the movement of the neck and projecting this into quantifiable posture data. This design is more visible than the previous two and may not be preferred by the user, since it is considered to be gender specific.

An analysis of the costs and the benefits for each of the three designs has been made and recorded in the Table shown in the next page. the undershirt strap scored the highest in terms of meeting our objectives and constraints. This design enables the use of several sensors and enables the calculation of the thoracic angle directly from the thoracic spine movement.

The next step in the design cycle was component selection. We aimed to begin by presenting a proof of concept and experiment with the data collection. Thus, it was decided that it would be a good solution to start with a computing and sensing unit that encompasses plenty of features as to not limit the system. Therefore, the size at this stage was not put into consideration but rather the usefulness. The next section in this reports discusses the methods of obtaining the angles, theoretically at first, and then moves on to obtaining the angles using sensing technologies. More details on component selection and iterations is discussed in section 6 of this report.

|  | 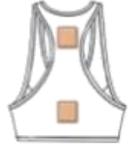 Design 1 | 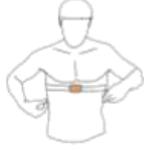 Design 2 | 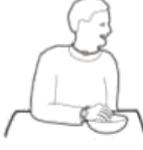 Design 3 |
|---|---|---|---|
| Objectives ||||
| 1) Safety | 1 | 1 | 1 |
| 2) Reliability | 1 | 0.5 | 0 |
| 3) wearability | 1 | 0.5 | 0.5 |
| 4) cost | 0.5 | 1 | 0 |
| Constraints ||||
| Power consumption must not exceed 800 mAh per 15 hours. | 0.5 | 1 | 0.5 |
| Size and thickness must not exceed 2.5 cm. Weight should not exceed 70 grams. | 0 | 0.5 | 1 |
| Can be used with various activities such as sports. (Versatile) | 1 | 0 | 0 |
| Should cover more than 3 vertebras of the spine | 1 | 0.5 | 0 |
| Error tolerance of ± 5% | 1 | 0.5 | 0 |
| Final Score ||||
|  | 7 | 5.5 | 3 |

# 4. THEORETICAL FOUNDATION

## 4.1 AHRS

An Attitude Heading and Reference System (AHRS) is a 3-axis system that is used to provide real-time 3D position which are the pitch, roll and yaw. using sensors with 3-axis magnetic, 3-axis acceleration, and 3-axis gyro. This system was developed to replace traditional gyroscopic instruments usually used in flight as they have better reliability and accuracy. The main difference between an inertial measurement unit (IMU) and a AHRS is the processing unit that is used to give the attitude as well as the heading information in the AHRS as opposed to the IMU which just gives sensory output to an external device that has to compute the attitude and heading independently. Additionally, AHRS are usually boarded with an extended Kalman filter [44].

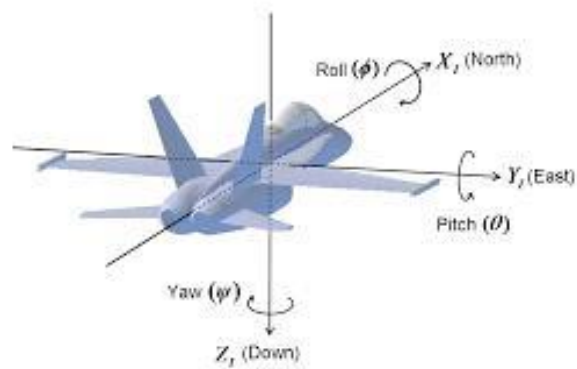

Figure 7: Attitude heading reference system

## 4.2 QUATERNIONS AND EULER ANGLES

Attitude and Heading Sensors are used to obtain both Euler angles as well as Quaternions. Euler angles are one of the several possible ways to represent rotational 3D orientation of an object. Euler angles are simple but are limited by the phenomenon known as the Gimbal Lock where the sensor orientation can't be expressed using Euler angles. This happens when the pitch angle is around 90 degrees. For this reason, quaternions, are used. Quaternions are another way of rotational representation, however, they consist of 4 elements, 3 complex values and one real value [45].

Quaternions provide an alternative measurement technique that does not suffer from gimbal lock. Quaternions are less intuitive than Euler Angles and the math can be a little more complicated. This is used to encode from the inertial rotation frame which is an is an Earth-fixed coordinate frame to the body-frame which is a frame that remains in alignment with the sensor [23].

To encode the rotation from the inertial frame to the body frame of the sensor we use the unit vector quaternion defined as $q_i^b$

$$q_i^b = (a\ b\ c\ d)^T$$

where T is the vector transpose operator. Here, the element a definer the rotation that should be done on the vector and is a scalar quantity. The remaining elements *b, c,* and *d* are the vector elements, that we perform the rotation on.

## 5. PRELIMINARY EXPLORATIONS

### 5.1 DATA ACQUISITON

**1) Testing data communication and types of data being sent.**

After purchasing and obtaining the sensor, the first things we did was to test the sensor and have a feel of what kind of data we are receiving. To do so, we connected the sensor to a PC and wrote a program that allows us to communicate with the sensor via a COM port. The sensor output data consisted of the IMU data for all 3 measurements of accelerator, gyroscope and magnetometer, each containing a triple axis. Moreover, the attitude consisting of the Euler angle was also given.

**2) Plotting the data against time**

We started by randomly and quickly moving the sensor in all directions and plotted the average value of each measurement against time. This plot shows the raw data obtained, prior to any form of calculation.

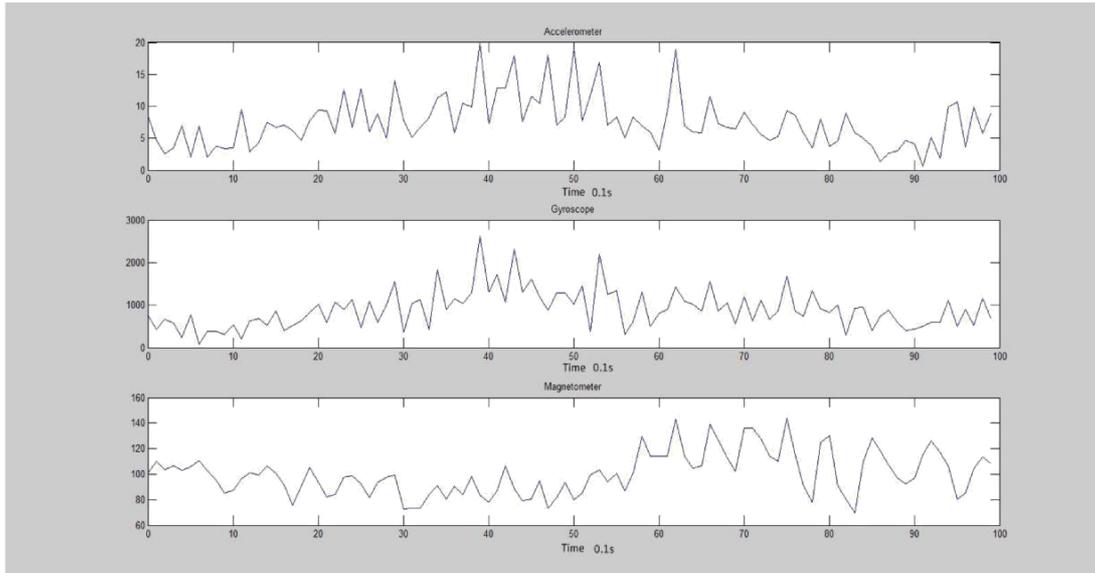

Figure 8: AHRS data graphing

**The next step is to obtain an angle and verify that a specific angle change in the sensors position reflects an equal angle change in the output data.**

## 5.2 ANGLE MEASUREMENT

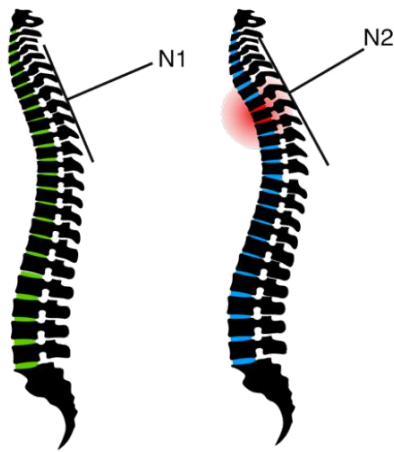

Figure 9: Location of sensor on spine

The angle was measured from the sensor data by first obtaining the direction cosine matrix (DCM) as shown below by using the values of the quaternions and substituting them into the matrix. Then the last 3 x 1 vector column from the DCM is extracted as it represents the normal vector to the sensor. The thoracic angle is found by using the dot product of the two normals from the calibrated upright position vector and the current position vector of the sensor. The magnitude of the new vector then gives us angle displacement. The graph below shows the variation of this angle with time as the user changes postural positions.

$$Quaternions = \begin{bmatrix}(\beta_0 & \beta_1 & \beta_2 & \beta_3)\end{bmatrix}$$

$$DCM = \begin{bmatrix} \beta_0^2 + \beta_1^2 - \beta_2^2 - \beta_3^2 & 2(\beta_1\beta_2 + \beta_0\beta_3) & 2(\beta_1\beta_3 - \beta_0\beta_2) \\ 2(\beta_1\beta_2 + \beta_0\beta_3) & \beta_0^2 - \beta_1^2 + \beta_2^2 - \beta_3^2 & 2(\beta_2\beta_3 + \beta_0\beta_1) \\ 2(\beta_1\beta_3 + \beta_0\beta_2) & 2(\beta_2\beta_3 - \beta_0\beta_1) & \beta_0^2 - \beta_1^2 - \beta_2^2 + \beta_3^2 \end{bmatrix} \quad N = \begin{bmatrix} 2(\beta_1\beta_3 - \beta_0\beta_2) \\ 2(\beta_2\beta_3 + \beta_0\beta_1) \\ \beta_0^2 - \beta_1^2 - \beta_2^2 + \beta_3^2 \end{bmatrix}$$

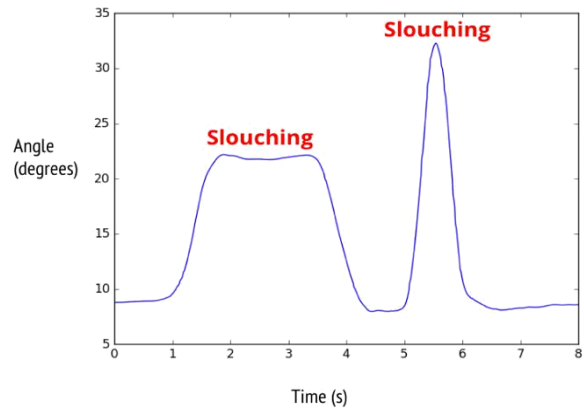
Figure 10: significant versus insignificant slouching

The graph above demonstrates a test case where the user slouches twice within 8 seconds. The first slouch was however longer, while the second one was due to sudden movement. Therefore, our system requires to differentiate between a significant and an insignificant slouch that will likely occur a lot throughout the day. The user should not be alerted unless the timespan of the slouch exceeds a specific critical value, such as 3secs. Therefore, a timer needs to be incorporated into the system for this purpose and initiated at the detection of a slouch.

## 5.3 PROBLEMS FACED WITH OBTAINING ANGLE

The two peaks shown represent a slouching position and a bending position. It is important to note that while slouching is considered poor posture, bending is not and is part of daily activities such as picking a pen up from the floor and many other similarly frequent movements. It will therefore be unfavorable for the device to go off every time the user bends. We should therefore be able to find a way to separate these two cases.

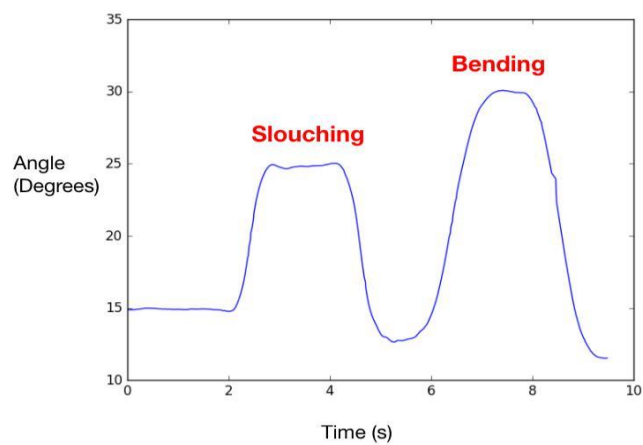
Figure 11: Bending versus slouching

In order to expand on the problem of discriminating between slouching and bending positions, new data was obtained and plotted from the IMU and then analyzed. Graphs for each of the acceleration, gyroscope and magnetometer reading were plotted separately in order to identify any possible distinctions that can help us to solve the problem of differentiating between the bending posture and the slouching posture.

The table below shows the postural stand for each time division. The main aim is to look into whether there exists any unique characteristics in the three graphs between the two time periods 100-240ms (slouching) and 350-520ms (bending) where the thoracic angle is of approximately the same value. For the Acceleration and Magnetometer graphs, we can see that those regions are almost identical, which tells us nothing useful.

However, in the gyroscope graph, the steepness of the value for the slouching position is much greater than that for the bending position. Therefore, we need to further investigate any possible distinctions that may be unseen in the graphs. This motivated us to go for a machine learning approach that will be discussed in further detail in the next section.

Table 2: Data for graphing

| Data Interpretation ||
|---|---|
| Time range (ms) | position |
| 0-100 | Upright |
| 100-240 | Slouching |
| 240-350 | Upright |
| 350-520 | Bending |
| 520-600 | Upright |

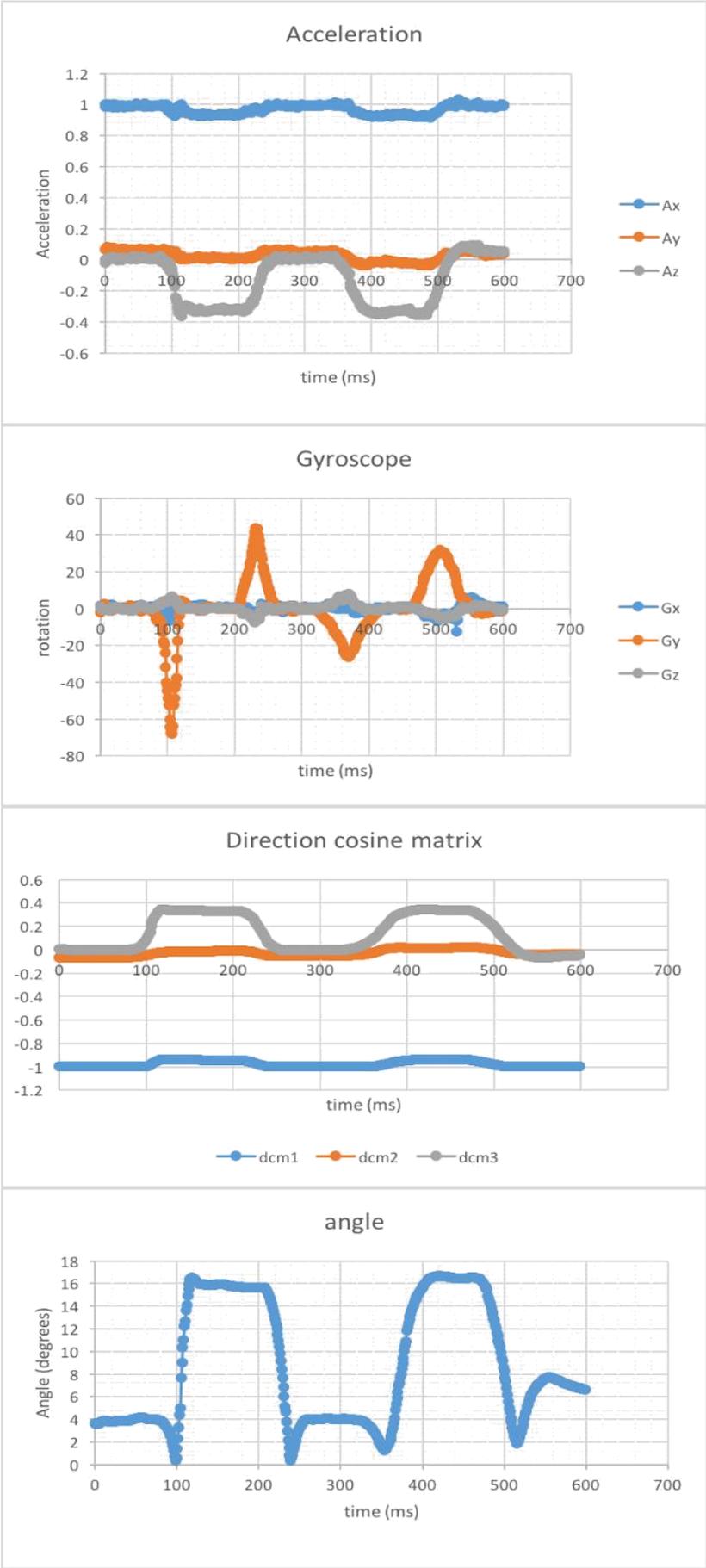

Figure 12: Data set graphs for machine learning

## 5.4 ATTRIBUTE SELECTION APPROACH

In order to investigate any hidden relationship between sensor data attributes and posture position, we created a database consisting of the data attributes and records from the previously discussed graphs and carried out a Principal component analysis (PCA) procedure that is based on the transformation of a data set into a set of linearly uncorrelated attributes known as principal components. An n-dimension ellipsoid is designed based on the data set, where its axis represents a principal component. The variance of the axis is then proportional to the axis length. The axes of the ellipsoid are found by calculating and then subtracting the mean value for each attribute. The covariance matrix is then calculated and the eigenvectors are obtained and normalized. Each eigenvector then represents an axis of the data. The variance of the attribute or the axis can be calculated by dividing its corresponding eigenvector value with the sum of all the eigenvectors.

After running the selection algorithm on the dataset containing all 9 degrees of freedom, in addition to quaternions and direction cosine matrix values, using Weka, a machine learning software, the following attributes were selected as having the highest variance and most distinct feature. The table below shows the eigenvectors and the ranking of the attributes, with accelerometer and gyroscope readings being selected as highest ranking features.

Table 3: Eigen vectors and ranking of attributes

| V1 | V2 | V3 | V4 | V5 | V6 | Attribute |
|---|---|---|---|---|---|---|
| -0.2983 | 0.0318 | 0.0095 | 0.0031 | -0.0283 | -0.0142 | Ax |
| -0.1891 | -0.2888 | 0.0489 | -0.0268 | 0.0075 | 0.0047 | Ay |
| -0.2938 | 0.116 | 0.0039 | 0.0061 | -0.0421 | -0.0097 | Az |
| 0.0231 | -0.0186 | 0.0196 | 0.7101 | 0.0461 | -0.7014 | Gx |
| -0.0825 | 0.1591 | -0.0401 | -0.0369 | 0.9818 | 0.02 | Gy |
| -0.0275 | -0.051 | 0.033 | 0.6998 | 0.0193 | 0.7109 | Gz |
| -0.2932 | 0.1228 | 0.0025 | 0.0067 | -0.0443 | -0.0087 | x |
| -0.2987 | 0.0244 | 0.0111 | 0.0018 | -0.0267 | -0.0105 | y |
| -0.2974 | 0.0631 | 0.0082 | 0.0038 | -0.0334 | -0.0095 | z |
| -0.189 | -0.475 | 0.047 | -0.0261 | 0.0699 | -0.0187 | w |
| 0.2986 | -0.0273 | -0.0101 | -0.0012 | 0.0283 | 0.0116 | dcm1 |
| 0.2881 | 0.1489 | -0.0224 | 0.0063 | -0.0045 | 0.0184 | dcm2 |
| 0.2939 | -0.1154 | -0.0036 | -0.0066 | 0.0426 | 0.0084 | dcm3 |
| -0.0308 | -0.0646 | -0.7027 | 0.0237 | -0.0203 | 0.0041 | mag2 |
| -0.0308 | -0.0646 | -0.7027 | 0.0237 | -0.0203 | 0.0041 | mag3 |

This does not tell us anything different from what we already know. Accelerometer and gyroscope readings are used in calculating the angle and therefore there are inevitably going to rank the highest. This confirms that there are indeed not sufficient data attributes that can aid us in discriminating between the two postures of bending and slouching.

## 5.5 CONCLUSION OF PRELIMINARY EXPLORATIONS

The first attempt at obtaining the thoracic angle was successful in terms of the accuracy of the angle, however, some problems have arisen. The main problem is the inability to distinguish between the two cases of slouching, which is considered under poor posture, and bending, which is a normal postural position. This was mainly due to the inability of a single sensor to detect enough information to differentiate the two.

The data was analyzed by plotting graphs of every measurable quantity from the sensor, however, no visual clear distinctions could be found. Therefore, a machine learning approach was considered to attempt to uncover any hidden characteristics of significance. A principle component analysis (PCA) was done, but it gave a similar outcome to the visual analysis test. Therefore, we concluded that a single sensor is not sufficient to make this distinction and other sensors need to be added to our system. This will be discussed further in the next session.

# 6. SYSTEM OVERVIEW

As previously discussed, the issue with using one IMU gives inaccurate results in terms of differentiating between slouching and bending which also makes the wearable device inconvenience to the user due to the vibration of the motor in both cases.

In the second prototype, an IMU and a flex sensor were used to fix the problem with the first prototype which is the discrimination between bending and slouching. What will differentiate between bending and slouching is the bending of the flex sensor. When the angle is obtained from the IMU, the microcontroller checks if the flex sensor has bent. If there is a change in the resistance value of the flex sensor, then a high signal will be sent to the buzzer to vibrate. But if there was no change in the flex sensor resistance, then bending will be detected and the motor will not vibrate.

## 6.1 SYSTEM DESIGN

In this section, the main parts of the system and its functions are represented by the block diagram. Then, the flowchart clarifies the data flow in the system and the decisions made to control events.

### 6.1.1 BLOCK DIAGRAM

The postural monitoring system contains three main parts which are sensor unit, control unit and power unit as shown in figure 12 below.

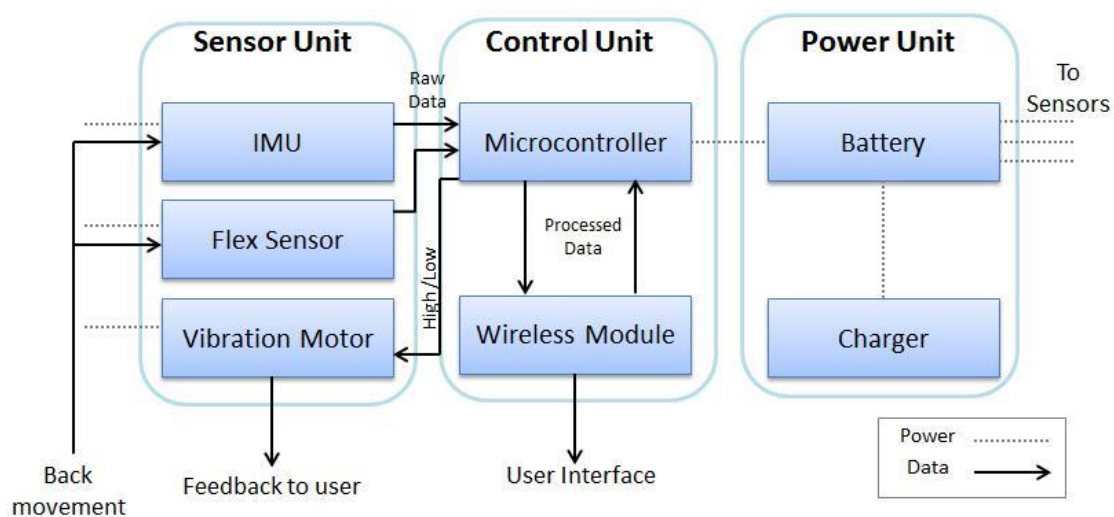

Figure 13: system block diagram

The sensor unit mainly collects data based on the user's movement of the back or more specifically, the spine. This data is then sent to the control unit for data processing. The control unit makes decisions based on the rotational data along with flex sensor data obtained from the sensor unit. Both sensing units and control units are powered by the power unit. It is responsible for supplying power to the whole system though a battery and a charger.

## 6.1.2 SYSTEM FLOW CHART

The data flow and the decisions taken by the microcontroller is explained in the figure below and it's followed by a description that will clarify each case in the chart.

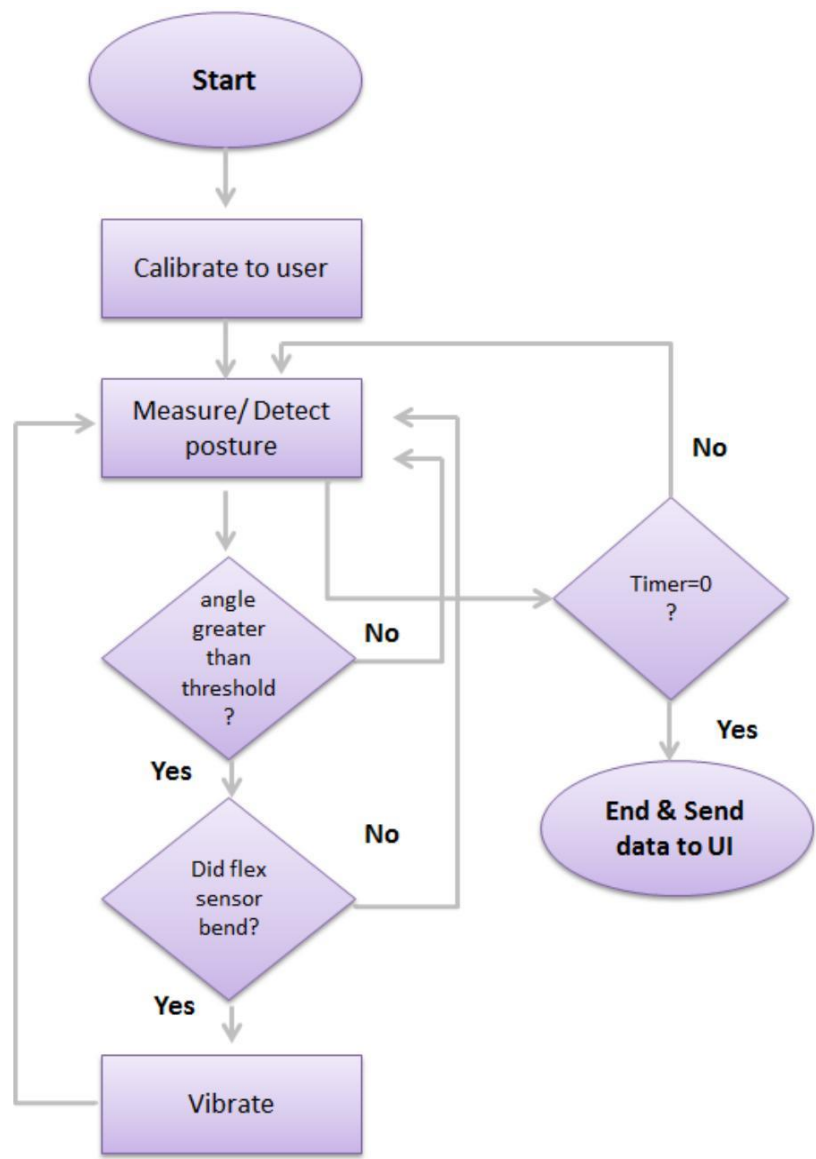

Figure 14: Flowchart of the Microcontroller

1) **Start** – Device is turned on.
2) **Calibrate to User** - The microcontroller will create a reference in the first 10 seconds the user wears the device and stands still for calibration.
3) **Measure/Detect Posture** – The angle is being calculated by the microcontroller.
4) **Send to Bluetooth** - The calculated angles will be sent to the Bluetooth to be sent to a user interface.
5) **Angle Greater Than Threshold?** - Microcontroller compares between the measured angle with the reference angle and based on that and the next case it will decide if the user is having a good posture.
6) **Did Flex Sensor Bend?** - Microcontroller checks if the flex sensor bent. If it did, then it means that the user is slouching, but if it didn't, it means that the user is bending.
7) **Vibrate** - If poor posture is detected (based on the previous two cases), the microcontroller sends a high signal to the buzzer which will cause it to vibrate to notify the user to adjust their posture.
8) **Timer = 0?** – This means that previously calibrated posture doesn't have to be maintained and the user can move.
9) **End & Send to User Interface** - The angular data will be sent to a use interface so the user can keep track with his progress.

## 6.2 HARDWARE SELECTION

This section lists the modules used in the system design and it includes the main three units which are the sensor unit, control unit and power unit as previously mentioned. The selection of the modules was made based on studies of the specs, availability of the components and experimental results.

### 6.2.1 SENSOR UNIT

**1) IMU (MY AHRS+)**

The sensor we will be using is the myAHRS+ sensor that consists of a high performing Attitude Heading Reference System (AHRS) with 9 degrees of freedom. This system has a triple axis 16-bit accelerometer, a triple axis 16-bit gyroscope, and a triple axis 13- bit magnetometer. This sensor was manufactured and designed by WITHROBOT Co [19]. I2C bus is used to pass information to the microcontroller.

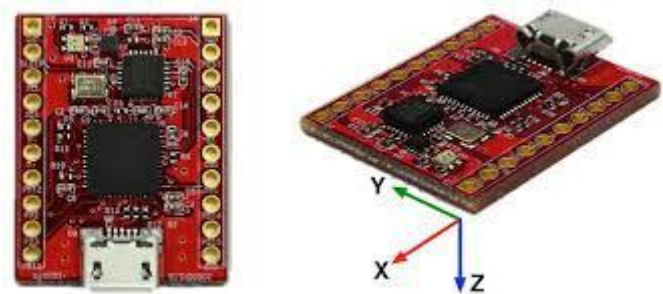

Figure 15: MyAHRS+ sensor [20]

- Triple axis 16-bit gyroscope: ± 2000 dps
- Triple axis 16-bit accelerometer: ± 16 g
- Triple axis 13-bit magnetometer: ± 1200 μT
- Extended Kalman filter
- 100 Hz frequency output rate
- Provides Attitude including Euler angle and Quaternion
- USB: Virtual COM PORT
- UART: Standard baud rates up to 460800 bps
- I2C: up to 1kHz

### 2) FLEX SENSOR

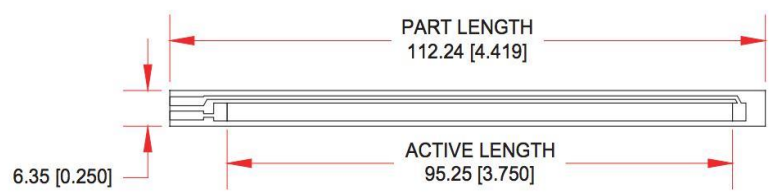

Figure 16: Flex Sensor [21]

A simple flex sensor is a resistive device that can detect bending through changes in the resistance across the sensor that is proportional to the amount of bending. The flex sensor is unidirectional as it detects bending in one direction, which is sufficient for our purposes. The flex sensor we will be using is the SEN-08606 4.5'' sensor. The output of the sensor is in volts and can be manipulated to display angle bending. The flex sensor is not sufficient to solely measure the thoracic angle of the back region as it is not sensitive enough for small angle changes which need to be detected by our system. It will therefore be used as a supportive sensor alongside the IMU sensor.

The specifications of the sensor are:

- Flat Resistance: 10KΩ-Resistance Tolerance: ±30%
- Bend Resistance Range: 60KΩ to 110KΩ
- Power Rating: 0.50 Watts continuous. 1 Watt Peak

### 3) VIBRATION MOTOR (LILYPAD VIBE BOARD)

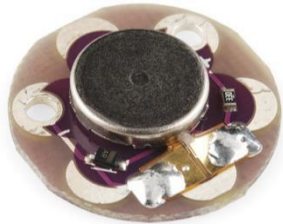

Figure 17: lilypad vibe board [21]

The lilypad vibe board is known as a great physical indicator because it notifies the user only not the surroundings. The motor on the top of the board is less likely to be damaged according to the manufacturer. In the system, the motor vibrates when poor posture is detected by receiving a high signal from a digital pin in the microcontroller.

## 6. 2.2 CONTROL UNIT

### 1) MICROCONTROLLER (ATMEGA328)

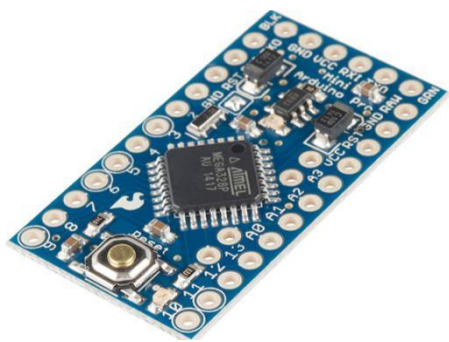

Figure 18: Arduino Pro Mini 3.3v [21]

The microcontroller receives data from the sensors via the I2C bus, processes it to compute the angles and checks if the user has exceeded the threshold value specified. If a poor posture is detected (which is done by comparing the output from the IMU and the bending of the flex sensor), the MCU sends a signal to the vibration motor to provide feedback to the user. The microcontroller also communicates wirelessly to a user interface to allow the user track his posture daily. The MCU board is really small and suits our needs, its dimensions are 0.7x1.3" (18x33mm) and it can be directly connected to the FTDI cable.

The main features of the board according to the manufacturer are:

- ATmega328 running at 8MHz with external resonator (0.5% tolerance)
- Low-voltage board needs no interfacing circuitry to popular 3.3V devices and modules (GPS, accelerometers, sensors, etc)
- 0.8mm Thin PCB
- Weighs less than 2 grams
- 3.3V regulator
- Over current protected
- DC input 3.3V up to 12V
- Analog Pins: 8
- Digital I/Os: 14

## 2) WIRELESS MODULE (BLUESMIRF BLUETOOTH MODULE)

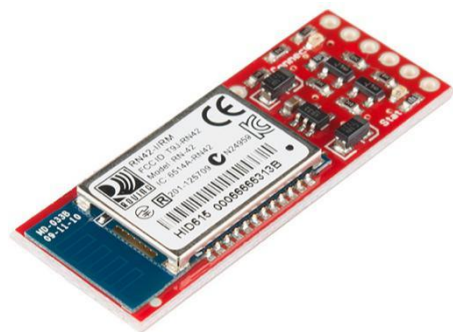

Figure 19: BlueSMiRF silver bluetooth modem [21]

The Bluetooth modem receives angular data and muscle flex data that has been transmitted from the microcontroller and relays them to a computer. Then, the output will be displayed on a computer forming a graph of user's posture during a day.

The specifications of the modem according to the manufacturer are:

- v6.15 Firmware
- FCC Approved Class 2 Bluetooth Radio Modem
- Extremely small radio - 0.15x0.6x1.9"

- Very robust link both in integrity and transmission distance (18m)
- Hardy frequency hopping scheme - operates in harsh RF environments like WiFi, 802.11g, and Zigbee
- Encrypted connection
- Frequency: 2.402~2.480 GHz
- Operating Voltage: 3.3V-6V
- Serial communications: 2400-115200bps
- Operating Temperature: -40 ~ +70C
- Built-in antenna

## 6. 2.3 POWER UNIT

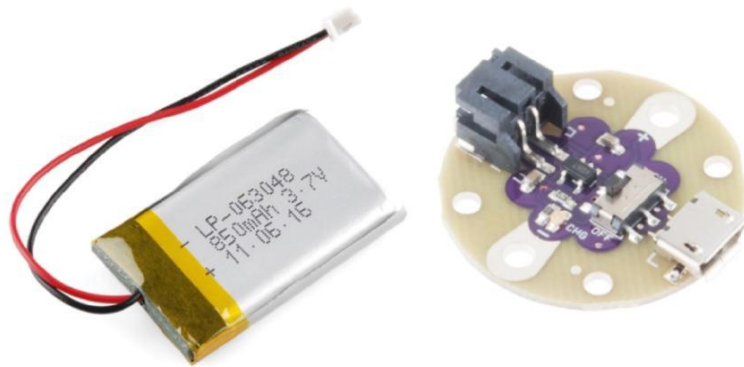

Figure 20:Lithium Ion Battery - 850mAh and LilyPad Simple Power

A rechargeable battery that outputs 3.7V at 850mAh is used to power the system. A LilyPad Simple Power charger board is used between the LiPo battery and the whole system. This board supplies regulated power to the system and enables the user to charge the Lithium Polymer by connecting it to a power source using a micro USB connection. The LilyPad Simple Power allows switching the lipo battery on or off, and it's provided by a JST connector.

## 6.3 FINAL SYSTEM DESIGN

After many design iterations and experimenting with different sensors, the figure below shows the final system and components used. However, in the figure below, the sensor shown is just an indicator for the myAHRS+ sensor as they use the same connection (i2c).

Three different sensors were used in the same system and each sensor was evaluated based on its size and data accuracy. The three sensors were MPU6050, Razor 9DOF and MyAHRS+. The MPU 6050 was very small in comparison with the other two, however it didn't not contain a magnetometer as compared to the other two. Using the same angle derivation method, we've made, MPU 6050 gave values that were very off. We tried compensating for the gyroscope drift as well as using various filtering methods to get more accurate quaternion data. However, most filtering techniques needed magnetometer data, and this was unavailable for the MPU6050. As for the Razor IMU, it was much larger than the myAHRS+ and had a less accuracy rate than myAHRS+ based on the system specs. Both were tried and myAHRS+ was more fitting with our system requirements and specifications.

### 6.3.1 SCHEMATIC DIAGRAM

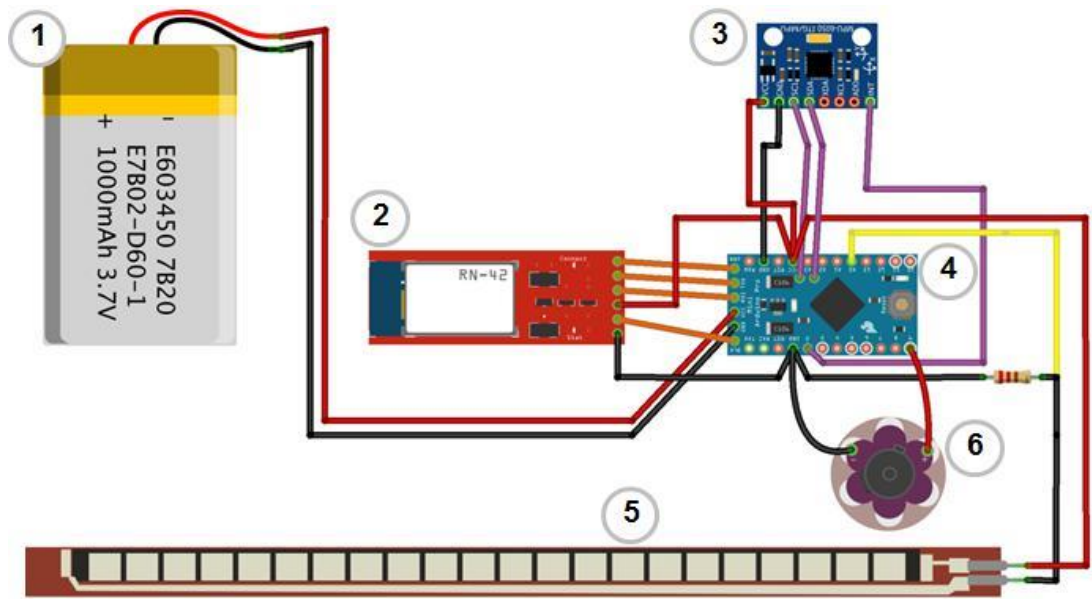

Figure 21: System schematic

**SCHEMATIC COMPONENTS**

1- **Rechargeable battery:** Lithium Ion Battery - 850mAh

2- **Bluetooth modem:** BlueSMiRF Silver

3- **IMU: MyAHRS+ sensor**

4- **Microcontroller:** Arduino Pro mini 328 - 3.3V/8MHz

5- **Flex sensor 4.5\"**

6- **Vibration motor:** LilyPad Vibe Board

First step, was to pin the whole system onto a prototyping breadboard for a more versatile and easy connection or component changes. Moreover, the Bluetooth mate can not load the Arduino program onto the Arduino, thus an FTDI wire was used to connect the Arduino Pro Mini MCU to the computer in order to load the program. And since both the Bluetooth module and the FTDI wire use the same pins on the Arduino, there had to be a lot of disconnections and connections in order to experiment with the design and program. This caused a design challenge as we had to delay transferring the components into the wearable system and solder components directly to each other because it will be much harder to disconnect and connect Bluetooth once the system is embedded onto the wearable strap. The figure below shows the system implemented on a breadboard for program prototyping and testing data received from each of the different sensors.

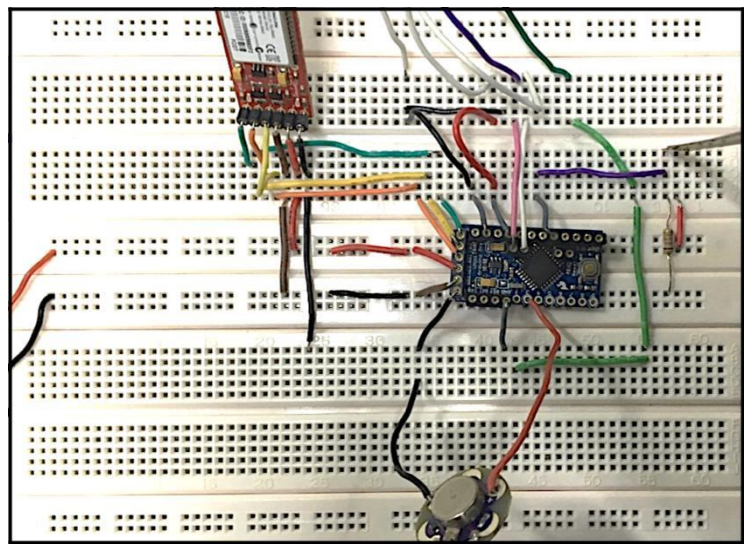

Figure 22: Prototype board

After finalizing the code and confirming that the vibration board vibrates only when both of the following events occur:

1) Flex sensor resistance exceeds the specified threshold (Resistance ≥ 33000 Ω).
2) Angle calculated using the data collected from the IMU sensor (Angle ≥ 20°).

All components were removed from the breadboard and were soldered directly to each other using wires in order to minimize the space it takes inside the wearable strap. The figure below shows the final system embedded inside the wearable strap. Figure 24 shows how the system is worn. The strap can also be worn beneath clothing for minimal disruptiveness to appearance. The next section covers the final testing of the system for usability and requirement verification.

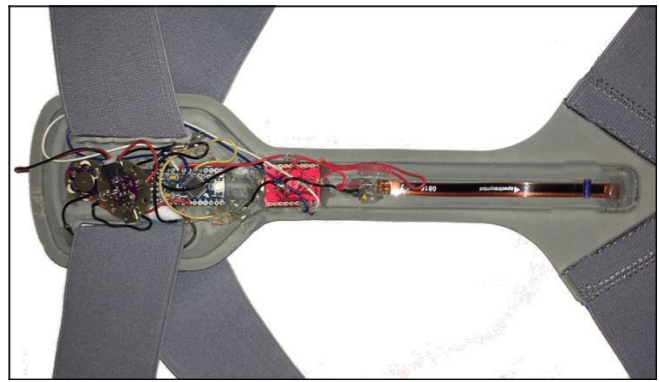

Figure 23: Wearable System close up

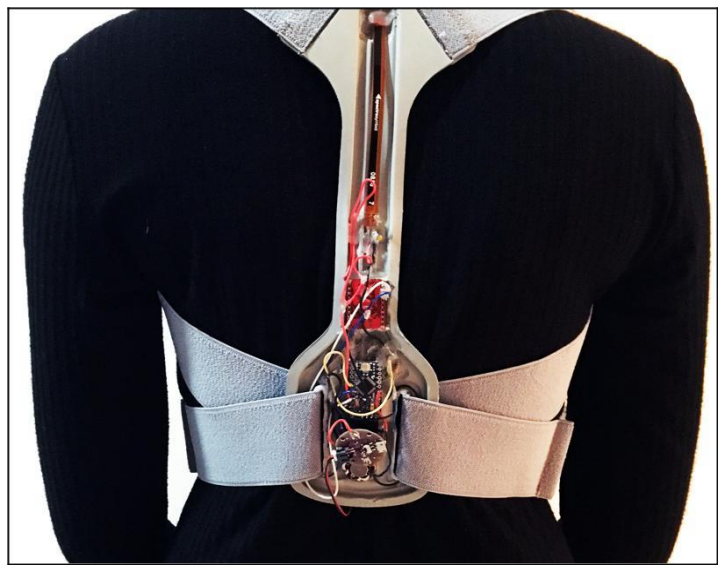

Figure 24: Final wearable system (Open top to show circuity)

## 7. TESTING AND RESULTS

The system was worn by one person for 2 continuous hours and was under examination the entire time. Every time the subject slouches or bends, it is noted down. The table of results below show different recorded cases. The first case is false positives that corresponds to the instances where the system detects a slouch, however the user is not slouching. True Positives on the other hand correspond to a real slouch detected by the system. Positives in general are the number of times the person has slouched, regardless of whether it was detected or undetected. Finally, False Negatives correspond to the times that the system failed to detect a slouch. Sensitivity is a measure of the true positive rate, or how many times the system was successful in detecting a slouch when it did happen. Our system had a sensitivity value of 85.1% which is a good number to begin with. In the future, several more testing phases need to be conducted on different ages, and gender groups to obtain a more reliable performance measure.

Table 4: testing results

| False Positive | 6 | False Negative | 8 |
|---|---|---|---|
| True Positive | 47 | Positives | 55 |
| | | | |
| Sensitivity | 85.1% | | |

## 8. FINANCIAL STATEMENT

The total cost of our system came up to be $589.375 which is around 2167 AED. This cost was spread across both semesters and the details of each purchase can be seen in the table below.

| Quantity | Name | Price (USD) |
|---|---|---|
| 1 | SparkFun FT231X Breakout | $14.95 |
| 1 | Arduino Pro Mini 328 - 3.3V/8MHz | $9.95 |
| 1 | LilyPad Arduino 328 Main Board | $19.95 |
| 1 | Lithium Ion Battery - 2000mAh | $12.95 |
| 1 | Lithium Ion Battery - 850mAh | $9.95 |
| 1 | Lithium Ion Battery - 1Ah | $9.95 |
| 1 | SparkFun Bluetooth Modem - BlueSMiRF Silver | $24.95 |
| 1 | LilyPad Vibe Board | $7.95 |
| 1 | LilyPad FTDI Basic Breakout - 5V | $14.95 |
| 1 | LilyPad Simple Power | $9.95 |
| 1 | Flex Sensor 4.5\" | $12.95 |

| 1 | LilyPad Arduino SimpleSnap | $29.95 |
| 2 | 9 Degrees of Freedom - Razor IMU | $74.95 |
| 2 | Conductive Thread Bobbin - 30ft (Stainless Steel) | $2.95 |
| 2 | Conductive Fabric - 12\"x13\" MedTex130 | $29.95 |
| 3 | myAHRS+ | $225 |
| 1 | Back Strap | $78 |
| | **Total** | **$589.375** |

# 9. CONCLUSION

In this paper we demonstrated a device for posture detection and correction via vibrational feedback. The system includes various hardware modules such as AHRS sensors, vibration motors, microcontroller and a wireless module. The data is firstly collected from the AHRS sensors that is positioned on thoracic part of the spine. Then, the data will be sent to the microcontroller in order to calculate the angular values that can verify whether the user's postural behavior by applying several mathematical operations. If poor posture has been detected, then the vibration motors will vibrate accordingly to notify the user. The wireless module will connect the control unit by another device/ an App. The mobile App will provide feedback to the user and display his daily posture routine. The main objective of this project was clarified that is to detect poor posture, train users to have a better posture until it becomes a habit and reduce chronic back pain that result from poor posture.

## 9.1 PROBLEMS AND CHALLENGES FACED

Initially, the main problem faced were understanding the physiological model of the spine and trying to find a way in which the curvature of the spine can be accurately measured. One issue was that we can not assume that the device is directly measuring the curvature of the spine because it is not directly attached to the spine and there are many other variables involved as it would differ from one person to the other. These variables include the density of the skin, the height of the person and the fact that one can not assume that everyone's spine has the same physical characteristics. Therefore, estimating the curvature of the spine was not simple. Moreover, finding a good design which will cover the specific part of the spine we were looking at was an issue. The reason being that the users will be varying in terms of height, weight and gender. That all had to be put in mind while designing the model.

Finally, moving the components from the prototype board and soldering them directly to each other for the final system design proved to be a challenge. As we had no prior experience with soldering components, there was always a risk of damaging the components. Since components take a long time to be delivered, we did not have the option of reordering. Some components were more sensitive than others which really added to the challenge of soldering the components together.

## 9.2 SYSTEM LIMITATIONS

Just as any system in the world, our system is not perfect and could be improved in many ways in the future. Our system has some limitations that need to be addressed in the future and these are listed below as follows:

- The current size of our system is larger than originally planned and this might cause some discomfort for the user as the components may be protruding. `This is mainly because of the large number of components that were used to build the system.

- During testing, we have noticed that the measured thoracic angle, although initially accurate, drifts by a small value with time. Therefore, after some time as the user continues to sit in an upright position, the system may detect a false positive.

- There is no way yet for the user to view their progress over time, or change the system settings such as sensitivity of angle.

- Our system focuses only on a single region of the spine, which is the thoracic region. Users might suffer from problems from other regions such as the kyphotic or neck area.

## 9.3 FUTURE WORK

This project was mainly focused on exploring the possible use of orientation sensors such as the AHRS in measuring postural changes in the sagittal plane of a human's body. There was a gap of knowledge in this area, as not many studies have gone deep into using orientation sensors in order to estimate the spinal curvature.

First, this system needs further verification and validation from experts in the medical field. Second, more work has to be invested into using more custom made components in order to shrink the size of the system for added comfort to the user and make it even less intrusive. Moreover, we must explore different ways where more areas of the spine can also be covered by the device. Either by using more flex sensors or more AHRS sensors.

In addition, the system could be expanded to add a user interface using a mobile application where users can track their day to day activities and progress and also be able to visually see their improvement.

It is important to note that the AHRS sensor gives us a lot of various readings, therefore can enable us to do various other things along with postural tracking. This means that the wearable system can also detect the user's activity such as running, walking, sitting or idle standing. This could help a lot with postural detection as it can vary according to the activity being performed. Moreover, the data received from the sensor can be used for fall detection and emergency situation detection, therefore it can be pretty useful for the elderly if developed accordingly.

# REFERENCES

[1] R. M. Shubair, "Robust adaptive beamforming using LMS algorithm with SMI initialization," in 2005 IEEE Antennas and Propagation Society International Symposium, vol. 4A, Jul. 2005, pp. 2–5 vol. 4A.

[2] R. M. Shubair and W. Jessmi, "Performance analysis of SMI adaptive beamforming arrays for smart antenna systems," in 2005 IEEE Antennas and Propagation Society International Symposium, vol. 1B, 2005, pp. 311–314 vol. 1B.

[3] F. A. Belhoul, R. M. Shubair, and M. E. Ai-Mualla, "Modelling and performance analysis of DOA estimation in adaptive signal processing arrays," in 10th IEEE International Conference on Electronics, Circuits and Systems, 2003. ICECS 2003. Proceedings of the 2003, vol. 1, Dec. 2003, pp. 340–343 Vol.1. 19

[4] R. M. Shubair and A. Al-Merri, "Robust algorithms for direction finding and adaptive beamforming: performance and optimization," in The 2004 47th Midwest Symposium on Circuits and Systems, 2004. MWSCAS '04, vol. 2, Jul. 2004, pp. II–589–II–592 vol.2.

[5] E. Al-Ardi, R. Shubair, and M. Al-Mualla, "Direction of arrival estimation in a multipath environment: An overview and a new contribution," in ACES, vol. 21, 2006.

[6] G. Nwalozie, V. Okorogu, S. Maduadichie, and A. Adenola, "A simple comparative evaluation of adaptive beam forming algorithms," International Journal of Engineering and Innovative Technology (IJEIT), vol. 2, no. 7, 2013.

[7] M. A. Al-Nuaimi, R. M. Shubair, and K. O. Al-Midfa, "Direction of arrival estimation in wireless mobile communications using minimum variance distortionless response," in Second International Conference on Innovations in Information Technology (IIT'05), 2005, pp. 1–5.

[8] M. Bakhar and D. P. Hunagund, "Eigen structure based direction of arrival estimation algorithms for smart antenna systems," IJCSNS International Journal of Computer Science and Network Security, vol. 9, no. 11, pp. 96–100, 2009.

[9] J. Samhan, R. Shubair, and M. Al-Qutayri, "Design and implementation of an adaptive smart antenna system," in Innovations in Information Technology, 2006, 2006, pp. 1–4.

[10]	M. S. Khan, A. D. Capobianco, S. M. Asif, D. E. Anagnostou, R. M. Shubair, and B. D. Braaten, "A Compact CSRR-Enabled UWB Diversity Antenna," IEEE Antennas and Wireless Propagation Letters, vol. 16, pp. 808–812, 2017.

[11]	R. M. Shubair and Y. L. Chow, "A closed-form solution of vertical dipole antennas above a dielectric half-space," IEEE Transactions on Antennas and Propagation, vol. 41, no. 12, pp. 1737–1741, Dec. 1993.

[12]	A. Omar and R. Shubair, "UWB coplanar waveguide-fed-coplanar strips spiral antenna," in 2016 10th European Conference on Antennas and Propagation (EuCAP), Apr. 2016, pp. 1–2.

[13]	R. M. Shubair and H. Elayan, "In vivo wireless body communications: State-of-the-art and future directions," in Antennas & Propagation Conference (LAPC), 2015 Loughborough. IEEE, 2015, pp. 1–5.

[14]	H. Elayan, R. M. Shubair, J. M. Jornet, and P. Johari, "Terahertz channel model and link budget analysis for intrabody nanoscale communication," IEEE transactions on nanobioscience, vol. 16, no. 6, pp. 491–503, 2017.

[15]     H. Elayan, R. M. Shubair, and A. Kiourti, "Wireless sensors for medical applications: Current status and future challenges," in Antennas and Propagation (EUCAP), 2017 11th European Conference on. IEEE, 2017, pp. 2478–2482.

[16]     H. Elayan and R. M. Shubair, "On channel characterization in human body communication for medical monitoring systems," in Antenna Technology and Applied Electromagnetics (ANTEM), 2016 17th International Symposium on. IEEE, 2016, pp. 1–2.

[17] H. Elayan, R. M. Shubair, A. Alomainy, and K. Yang, "In-vivo terahertz em channel characterization for nano-communications in wbans," in Antennas and Propagation (APSURSI), 2016 IEEE International Symposium on. IEEE, 2016, pp. 979–980.

[18]     H. Elayan, R. M. Shubair, and J. M. Jornet, "Bio-electromagnetic thz propagation modeling for in-vivo wireless nanosensor networks," in Antennas and Propagation (EUCAP), 2017 11th European Conference on. IEEE, 2017, pp. 426–430. 21

[19]     H. Elayan, C. Stefanini, R. M. Shubair, and J. M. Jornet, "End-to-end noise model for intra-body terahertz nanoscale communication," IEEE Transactions on NanoBioscience, 2018.

[20]     H. Elayan, P. Johari, R. M. Shubair, and J. M. Jornet, "Photothermal modeling and analysis of intrabody terahertz nanoscale communication," IEEE transactions on nanobioscience, vol. 16, no. 8, pp. 755–763, 2017.

[21]     H. Elayan, R. M. Shubair, J. M. Jornet, and R. Mittra, "Multi-layer intrabody terahertz wave propagation model for nanobiosensing applications," Nano Communication Networks, vol. 14, pp. 9–15, 2017.

[22]     H. Elayan, R. M. Shubair, and N. Almoosa, "In vivo communication in wireless body area networks," in Information Innovation Technology in Smart Cities. Springer, 2018, pp. 273–287.

[23]     M. O. AlNabooda, R. M. Shubair, N. R. Rishani, and G. Aldabbagh, "Terahertz spectroscopy and imaging for the detection and identification of illicit drugs," in Sensors Networks Smart and Emerging Technologies (SENSET), 2017, 2017, pp. 1–4.

[24]     L. Straker, C. Pollock and J. Mangharam, "The effect of shoulder posture on performance, discomfort and muscle fatigue whilst working on a visual display unit", *International Journal of Industrial Ergonomics*, vol. 20, no. 1, pp. 1-10, 1997.

[25]     A. Reeve and A. Dilley, "Effects of posture on the thickness of transversus abdominis in pain-free subjects",
*Manual Therapy*, vol. 14, no. 6, pp. 679-684, 2009.

[26]     *Back Pain Facts & Statistics*, 1st ed. Virginia: American Chiropractic Association.

[27]     *In Project Briefs: Back Pain Patient Outcomes Assessment Team (BOAT). In MEDTEP Update*, vol. 1 no. 1, Rockville, Agency for Health Care Policy and Research.

[28] Emirates 24/7, "62% of young people suffer lower back pain", 2015.

[29]     J. Riskind and C. Gotay, "Physical posture: Could it have regulatory or feedback effects on motivation and emotion?", *Motivation and Emotion*, vol. 6, no. 3, pp. 273-298, 1982.

[30]     L. Bey and M. Hamilton, "Suppression of skeletal muscle lipoprotein lipase activity during physical inactivity: a molecular reason to maintain daily low-intensity activity", *The Journal of Physiology*, vol. 551, no. 2, pp. 673-682, 2003.

[31]     KnowHowMD, *Thoracic curve*. http://www.knowhowmd.com/spine/disorders/kyphosis

[32]     A. Lindegard, C. Karlberg, E. Wigaeus Tornqvist, A. Toomingas and M. Hagberg, "Concordance between VDU-users' ratings of comfort and perceived exertion with experts' observations of workplace layout and working postures". *Applied Ergonomics* 2005; 36: 319-325

[33]  Dunne, L. E., Walsh, P., Hermann, S., Smyth, B., and Caulfield, B. Wearable monitoring of seated spinal posture. Biomedical Circuits and Systems, IEEE Transactions on 2, 2 (2008), 97–105.

[34] L. Dunne, P. Walsh, B. Smyth, and B. Caulfield. Design and evaluation of a wearable optical sensor for monitoring seated spinal posture. In Proc. of ISWC, Montreux, 2006. 5

[35] L. Dunne, S. Brady, B. Smyth, and D. Diamond. Initial devel- opment and testing of a novel foam-based pressure sensor for wearable sensing. J NeuroEngineering Rehabil, 2(4), 2005. 5,

[36]  G. A. Hansson, P. Asterland, N. G. Holmer, and S. Skerfving. Validity and reliability of triaxial accelerometers for inclinometry in posture analysis. Med Biol Eng Comput, 39(4):405–13, 2001. 4

[37]  K. Van Laerhoven and O. Cakmakci. What shall we teach our pants? In Proc. of ISWC, pp. 77–83, 2000. URL citeseer.ist.psu. edu/vanlaerhoven00what.html. 4

[38] W. Lin, M. Lee and W. Chou, "The design and development of a wearable posture monitoring vest", *2014 IEEE International Conference on Consumer Electronics (ICCE)*, 2014.

[39]     "Lumo Lift - Workrite Ergonomics", *Workrite Ergonomics*, 2016. [Online]. Available: http://workriteergo.com/lumo-lift/. [Accessed: 27- Aug- 2016].

[40] Q. Wang, W. Chen, A. Timmermans, C. Karachristos, J. Martens and P. Markopoulos, "Smart Rehabilitation Garment for posture monitoring", *2015 37th Annual International Conference of the IEEE Engineering in Medicine and Biology Society (EMBC)*, 2015.

[41] D. Giansanti, V. Macellari, G. Maccioni and A. Cappozzo, "Is it feasible to reconstruct body segment 3-D position and orientation using accelerometric data?", *IEEE Trans. Biomed. Eng.*, vol. 50, no. 4, pp. 476-483, Apr. 2003.

[42] "myAHRS+ (Attitude Heading Reference System) [0054]", *Ameridroid.com*, 2016. [Online]. Available: http://ameridroid.com/products/myahrs-attitude-heading-reference-system. [Accessed: 23- Aug- 2016].

[43]     "[myAHRS+] Introduction | WITHROBOT", *Withrobot.com*, 2016. [Online]. Available: http://www.withrobot.com/myahrs_plus_en/. [Accessed: 24- Aug- 2016].

[44] "What is an AHRS? - Sparton", *Sparton*, 2015. [Online]. Available: http://www.spartonnavex.com/ahrs/.

[Accessed: 25- Aug- 2016].

[45]"Understanding Euler Angles | CH Robotics", *Chrobotics.com*, 2016. [Online]. Available: http://www.chrobotics.com/library/understanding-euler-angles. [Accessed: 25- Aug- 2016].

[46]"Understanding Quaternions | CH Robotics", *Chrobotics.com*, 2016. [Online]. Available: http://www.chrobotics.com/library/understanding-quaternions. [Accessed: 25- Aug- 2016].

PROJECT TIMELINE The Gantt chart below covers the SDP II work.

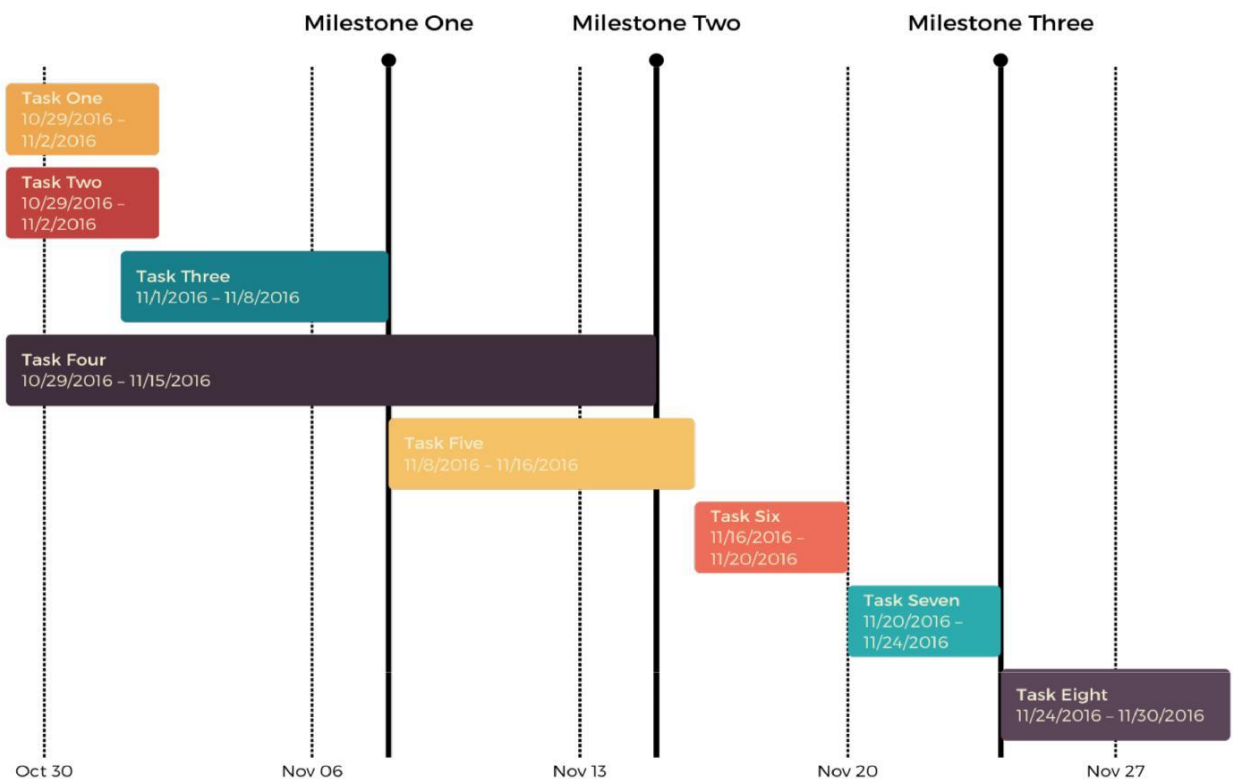

Figure 25: Fall 2016 Gantt chart

1. Set up sensor to MCU connection and program
2. Set up bluetooth connection
3. Set up all sensors and actuators to MCU
4. Embed in clothing
5. Transfer data to and from computer wirelessly
6. Set up feedback system
7. Test and validate
8. Finalize prototype, design and operation.

The Gantt chart below covers the work done during Spring 2015 semester for SDP I.

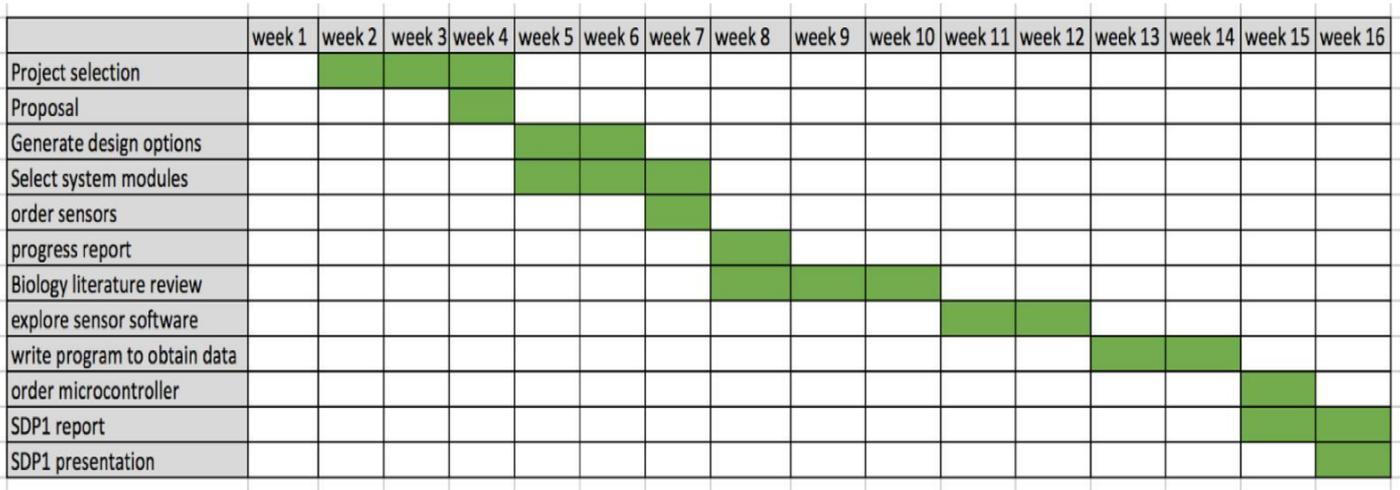

Figure 26: Spring 2015 Gantt chart.